# Chemical vapor deposition of carbon nanotubes on monolayer graphene substrates: reduced etching via suppressed catalytic hydrogenation using $C_2H_4$

*Kitu Kumar[1], Youn-Su Kim[1], Xin Li[1], Junjun Ding[1], Frank T. Fisher[1], Eui-Hyeok Yang[1]\**

Department of Mechanical Engineering, Stevens Institute of Technology, Castle Point-on-Hudson, Hoboken, NJ 07030 USA

eyang@stevens.edu

**In most envisioned applications, full utilization of a graphene-carbon nanotube (CNT) construct requires maintaining the integrity of the graphene layer during the CNT growth step. In this work, we exhibit an approach towards controlled CNT growth atop graphene substrates where the reaction equilibrium between source hydrocarbon decomposition and carbon saturation into and precipitation from the catalyst nanoparticles shifts towards CNT growth rather than graphene consumption. By utilizing $C_2H_4$ feedstock, we demonstrate that the low temperature growth permissible with this gas suppresses undesirable catalytic hydrogenation and dramatically reduces the etching of the graphene layer to exhibit graphene-CNT hybrids with continuous, undamaged structures.**

Recent efforts in fabricating three dimensional (3D) composite nanostructures consisting of two dimensional (2D) graphene and one dimensional (1D) nanomaterials of carbon[1-3] and conducting polymers[4] are of interest for a number of applications, including next-generation, high capacity, fast-discharge supercapacitors. For these types of energy storage applications, the advantages of graphene, such as large surface area-to-volume ratio and excellent conductivity, may be compromised due to self-aggregation resulting in poor charge transfer between the graphene flakes, the 1D materials, and the current collector. The growth of 1D nanostructures such as carbon nanofibers[5,6] or nanotubes[7,8] directly on graphene to yield hybrid 3D nano-architectures would, by design, circumvent this self-aggregation, while maintaining low contact resistance to enable effective electron transfer.[7-9] In our previous work, CNTs were grown by chemical vapor deposition (CVD) directly on graphene using $CH_4$ gas as a carbon source, and the performance of the resulting 3D nanoarchitecture as an advanced electrical double layer capacitor was characterized.[9] However, during this growth, the graphene layer was often found to be etched away at so-called "etched pits". The formation of these pits proceeded from hydrogenation[10-12] at 800°C in the presence of nickel (Ni) catalyst nanoparticles $((Ni)_{nanoparticle} + C_{graphene} + 2H_2 \rightarrow Ni + CH_4)$.[13] Elongated etched lines in the graphene are attributed to etching by mobile nanoparticles. Subsequently, the addition of $H_2$ from the catalytic decomposition of the carbon source during the CNT growth step further contributes to the etching effect and can fully remove the graphene substrate. This etching process of the graphene substrate during CNT growth has thus far not been studied in the literature.

Here we show that the high hydrocarbon conversion rate of $C_2H_4$, at lower temperature than $CH_4$[14] used in our previous study,[9] allows for an approach to CNT growth atop graphene substrates through fine tuning the process parameters including growth temperature and seed density. We confirm that the controlled use of $C_2H_4$ is essential for balancing the competing processes of carbon deposition and carbon removal, which ultimately block undesired etching of the graphene substrate during the CNT growth process.

**Results and Discussion**

After graphene-CNT structures were fabricated (Scheme 1), scanning electron microscopy (SEM), Raman spectroscopy, and transmission electron microscopy (TEM) imaging were conducted to characterize the structure of the CNTs grown on graphene. Of fundamental importance was confirming the growth of 1) graphene on the substrate and 2) CNTs on the graphene. TEM images of graphene, the graphene-CNT interface, and CNTs were analyzed (Figure 1) to assess the quality of the grown samples. Graphene was found to be monolayer in the majority of measured regions (Figure S1). Multiwalled CNTs were found to have a root in the graphene lattice as evidenced by Figure 1a. Figure

1b shows a CNT clearly growing out of the graphene layer. The ohmic contact between CNT and graphene, formed during such a growth,[15] is a necessity to facilitate charge transfer[16] between the two materials for energy storage applications. Additionally, the graphene planes of the multiwalled CNTs run parallel to the growth axis of the tube, as shown in Figure 1c, confirm that CNTs were grown in this process and not carbon nanofibers. For comparison, TEM images of a carbon nanofiber show a characteristic fishbone arrangement, as in Figure S2.

Raman spectra were taken at an excitation wavelength of 532 nm to assess the crystalline quality of the graphene-CNT structure. As shown in Figure 2a, the original as-grown graphene films exhibit the three distinctive peaks of $sp^2$ carbon, namely, D, G and G', corresponding to defects, $E_{2g}$ vibrations of the $sp^2$ bonds, and a second-order double-resonance process distinctive in graphene, respectively.[17] The G' band peak of the graphene shows a higher peak intensity than the G band with an intensity ratio $I_{G'}/I_G$ of 1.57 and can be fitted to a sharp, symmetric Lorentzian with a full width at half maximum (FWHM) of 31.8 $cm^{-1}$. The D band is minimal with an intensity ratio $I_D/I_G \sim$ 0.11, attributed to the presence of grain boundaries.[18] These characteristics strongly suggest that the grown graphene is high quality and monolayer.[19] In comparison, the Raman spectrum of the graphene-CNT hybrid structure presented in Figure 2b shows a broad G band, a suppressed 2D band, and a large defect peak of D band at ~ 1380 $cm^{-1}$ ($I_D/I_G \sim$ 0.50) arising from the addition of the grown CNTs. The G band FWHM increases from 17.2 $cm^{-1}$ in the graphene case to 39.8 $cm^{-1}$ in the graphene-CNT case, and is conspicuously split into two components, $G^-$ (1614.8 $cm^{-1}$) and $G^+$ (1574.6 $cm^{-1}$). The $G^-$ component arises from vibrations along the circumferential direction of the CNTs while the $G^+$ component arises from vibrations along the tube growth axis.[20] Further, this splitting exhibits a smearing effect due to the presence of multiple diameter graphene walls in the CNT which gives rise to the asymmetrical shape of the G band.[21] It is noteworthy that no radial breathing mode was observed in the graphene-CNT sample, due to this smearing effect.[22] These results, along with TEM characterization, confirm the successful growth of monolayer graphene followed by the growth of CNTs on the graphene substrate.

In order to study the impact of processing conditions on graphene etching during CVD, we first analyzed the impact of temperature during catalyst nanoparticle generation. In Figure 3a, a 3 nm Ni film atop graphene was exposed to only $Ar/H_2$ (400/50 sccm) flow at 800°C. This high temperature process dewetted[23-25] the Ni film, forming catalyst nanoparticles with highly variant size (Figure 3c) and low density. In this condition, which was intentionally created to detail the graphene etching phenomena, etched pits appeared in graphene around the nanoparticles (Figure 3a, white arrows). At this temperature, catalytic hydrogenation occurs wherein the carbon atoms in graphene enter the molten Ni droplet and subsequently react with $H_2$ at the surface of the droplet forming $CH_4$ gaseous species.[11,26] The two principal particle sizes observed in Figure 3a,c are attributed to a surface diffusion-based Ostwald ripening process.[27,28] The presence of certain etched lines (Figure 3a, blue arrows) appear to be caused by etching from mobile nanoparticles, which continues until the energetics for the motion reaction cease. These etched pits and lines are capable of originating at this temperature at any high energy, defectives sites in graphene such as grain boundaries. Indeed, the initiation of nanoparticle motion has been shown to occur due to attractive forces between the particle and carbon atoms with dangling bonds, rather than in the basal plane.[29]

We then analyzed the catalyst nanoparticle generation of a 3 nm Ni film atop graphene at 700°C under only $Ar/H_2$ (400/50 sccm). At this temperature, in sharp contrast with the 800°C data point, the Ni nanoparticles had a higher density on the graphene sheet (Figure 3b) with smaller diameter variation (Figure 3c) and the only hydrogen reaction active is hydrogen reduction[30] which produces even, circular cross-section nanoparticles at a high density (Figure S3). Without the presence of the hydrogenation effect or surface diffusion of the nanoparticles, damage to the graphene sheet was observed to be negligible. It is important to note here that in addition to the benefit of unetched graphene, well-shaped, high density catalyst nanoparticles lead to vertical, high density CNTs due to van der Waal attractions, thus providing more active surface area for envisioned energy storage applications.[9]

Having established the effect of temperature on the integrity of the graphene substrate during catalyst nanoparticle formation, we next analyzed the impact of CNT growth conditions on graphene etching. In Figure 4a, catalyst nanoparticle generation and

CNT growth was accomplished at a reaction temperature of 800°C. CNTs with relatively low density (approx. 2.8 × 10$^9$ cm$^{-2}$) were grown (Figure 4b). Observation of Ni nanoparticles near the top-most layers strongly suggests top-down growth, thus confirming the mobility of Ni nanoparticles from the thermal treatment step. The bottom-side of the graphene-CNT sample in Figure 4c (i.e., the side contacting SiO$_2$) shows a dramatic change in both the morphology of CNTs and the extent of graphene etching during the CNT growth. The graphene layer appears to have been fully etched away; the etched sites formed during catalyst nanoparticle generation are enlarged by hydrogen etching[31] from excess H$_2$ generated by the decomposition of C$_2$H$_4$. Additionally, the diameters of CNTs near the bottom-side are larger (38±13 nm), whereas growth away from the graphene layer produced smaller diameter CNTs (9.2±1.5 nm) This evidence suggests two stages of CNT growth rate; one stage proceeding from both carbon feedstock and graphene as the carbon source via etching and the second stage based solely on the carbon feedstock gas as carbon availability from the graphene diminished and the graphene layer was completely etched.

For comparison, consider Figure 5 wherein catalyst nanoparticles were generated at 700°C, but CNT growth was performed at 800°C. The mean CNT diameter near the graphene substrate is 25±7 nm. Here, we observed only partial etching of the graphene layer. The Ni nanoparticles are still present near the graphene substrate and the circular shape of the etched tracks at the point of contact indicate that the particles are largely non-mobile during the growth step, and further, that the additional H$_2$ from C$_2$H$_4$, which specifically decomposes near the catalyst nanoparticle, contributes to the hydrogenation reaction that can only occur at high temperature. We now discuss the difference between the 700°C and 800°C reactions.

The catalytic hydrogenation reaction tends to increase at a high temperature (800°C) where the density of the catalyst nanoparticles is low,[10-12] since the reaction equilibrium between source hydrocarbon decomposition and carbon saturation into and precipitation from the catalyst nanoparticles shifts towards consumption of graphene at the nanoparticle-graphene contact interface.[12] In general, as the Ni catalysts become supersaturated with carbon (from thermally decomposed hydrocarbon gas) towards CNT growth, the hydrogenation process limits the supersaturation state by removing carbon on the surface of the catalyst.[10-12] However, in the case of etching, since the Ni catalyst density is low, this carbon removal extends to the graphene layer. Therefore the graphene layer is etched via hydrogenation in a concentrated hydrogen environment supplied both by an $H_2$ source and the decomposed hydrocarbon source gases (such as $CH_4$ or $C_2H_4$) at high temperature. The catalyst generation process would then yield etched sites in graphene, which would become further enlarged during the CNT growth process, especially at 800°C.

In clear contrast, Figure 6 shows the successful fabrication of CNTs on graphene at 700°C. In Figure 6a, a high areal density of CNTs with mean diameter of 24±3 nm (Figure 6b) were directly grown on the graphene substrate without prominent signs of graphene etching. Similar CNT diameters were produced in Figure 5b since catalyst generation in both cases was at 700°C, with a larger standard deviation since some of the carbon source originated from the graphene. The successful result in Figure 6 is attributed to the lower process temperature of 700°C, which produces highly dense Ni catalyst nanoparticles and curbs etching of the graphene substrate (Figure 6c) during catalyst generation and CNT growth. Such conditions cannot be obtained at 800°C or higher, temperatures which are necessary to grow CNTs when using $CH_4$ as source gas (Figure S4).[14] Therefore, the process conditions selected for CVD growth of CNTs directly on graphene, such as catalyst nanoparticle density and the type of hydrocarbon gas, which impacts growth temperature and concentration of carbon and hydrogen in the reactor, should be carefully tuned to reduce the incidence of graphene etching by suppressing hydrogenation.

In summary, we have studied the graphene etching phenomena occurring during direct growth of CNTs on graphene, and have identified an approach to reduce etching of the graphene substrate by using $C_2H_4$ gas. We have shown that at high temperatures (800°C) a catalytic hydrogenation reaction results in the primary growth of etched pits and lines on the graphene layer followed by expansion of the etched sites via excess hydrogen during the CNT growth process. By using $C_2H_4$ gas as a hydrocarbon source for CNT growth under low temperature (700°C) and controlled gas ratio conditions, the catalytic hydrogenation reaction was dramatically suppressed to avoid etching of graphene during the CNT growth process. The successful fabrication of graphene-CNT structures has exceptional implications in applications where the continuity and integrity of the graphene layer is preserved.

**Experimental Section**

Large area graphene layers were grown on Cu foil (99.99% purity) by Atmospheric Pressure Chemical Vapor Deposition (APCVD).[9,32] A rolled Cu foil (13x60 cm$^2$) was placed at the center of a 2-inch quartz tube in a horizontal 3 zone CVD reactor and heated to 1000°C under flow of H$_2$ and Ar. A high temperature annealing step was carried out to increase the grain size of the Cu foil, ensuring high quality graphene films. During the growth step, CH$_4$, H$_2$ and Ar were fed through the system at flow rates of 50, 15 and 1000 sccm, respectively, for 4 min. Subsequently, the sample was rapidly cooled to room temperature by flowing pressurized H$_2$ and Ar gases in the furnace. Thermal tape was then attached to the graphene/Cu stack and a force of 9.8 N-cm$^{-2}$ was applied to ensure adhesion between the tape and graphene. The Cu foil was fully etched using citric acid Cu etchant (Transene, Inc), followed by several deionized (DI) water baths to remove residual etchant. The tape/graphene stack was transferred to a cleaned SiO$_2$ wafer (4 inch) and uniform force was applied for 10 minutes. The substrate was heated to 123°C to detach the thermal tape and any remaining adhesive residue was removed with boiling acetone (90°C) and DI water.

After this transfer, approximately 3 nm of Ni catalyst film was deposited on the graphene layer via electron beam deposition using an Explorer 14 (Denton Vacuum) PVD evaporator. The catalyst/graphene sample was then placed in the CVD reactor and heated to the desired growth temperature while flowing Ar gas in preparation for the introduction of the reaction source gas. The sample was held for 30-45 min at the desired temperature to ensure the thermal breakdown of the Ni thin film into catalyst nanoparticles. After this step, if necessary, the reaction temperature was adjusted, and a mixture of C$_2$H$_4$ (99.97%, GTS Welco), H$_2$ and Ar gases was flown through the furnace for CNT growth on graphene via the vapor-liquid-sold (VLS) process within the Ni catalysts. After growth was completed, the tube was cooled down to room temperature under Ar flow (400 sccm) only.

For TEM imaging, an as-grown graphene-CNT sample was sonicated in ethanol to get a low-density, uniform solution of the material. A microdropper was used to drop this solution on a lacey carbon TEM grid. Low-magnification TEM images of likely graphene-CNT areas were taken using an FEI CM20 field-emission S/TEM with 200kV. voltage. After pinpointing such regions, high magnification TEM images were taken using a JEOL JEM2100F Transmission Electron Microscope at 200kV.

**Acknowledgements**

This work was supported by the Robert Crooks Stanley Fellowship and by National Science Foundation (DMR-0922522, EECS-1040007, ECCS-1104870, and EEC-1138244) and Air Force Office for Scientific Research (FA9550-11-1-0272, FA9550-12-1-0326).

**Supporting Information**

TEM images of graphene and carbon nanofibers; AFM images of graphene and catalyst nanoparticles; SEM images of $CH_4$ growth of CNTs; optical transmission spectra and Raman spectra of graphene and graphene-CNT samples. This material is available free of charge via the Internet at http://pubs.acs.org.

**Schemes and Figures**

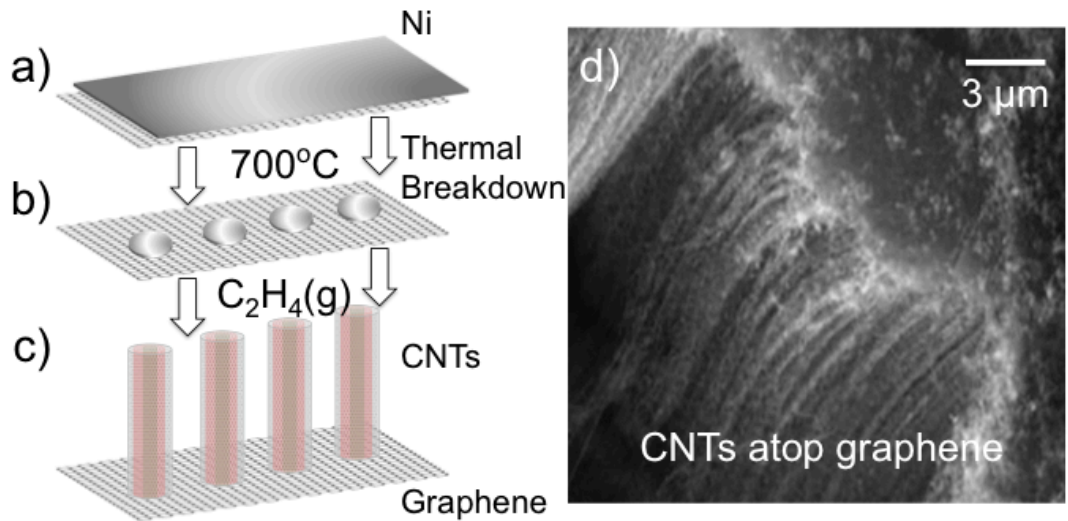

**Scheme 1.** CNT growth process on a graphene substrate. a) Graphene with 3 nm of electron beam evaporated Ni film b) Thermal treatment of the 3 nm Ni film to form Ni nanoparticles. c) CNT growth from Ni catalyst nanoparticles. d) SEM image of vertically grown CNTs atop a graphene substrate. Sample is intentionally broken and peeled off for purpose of observation.

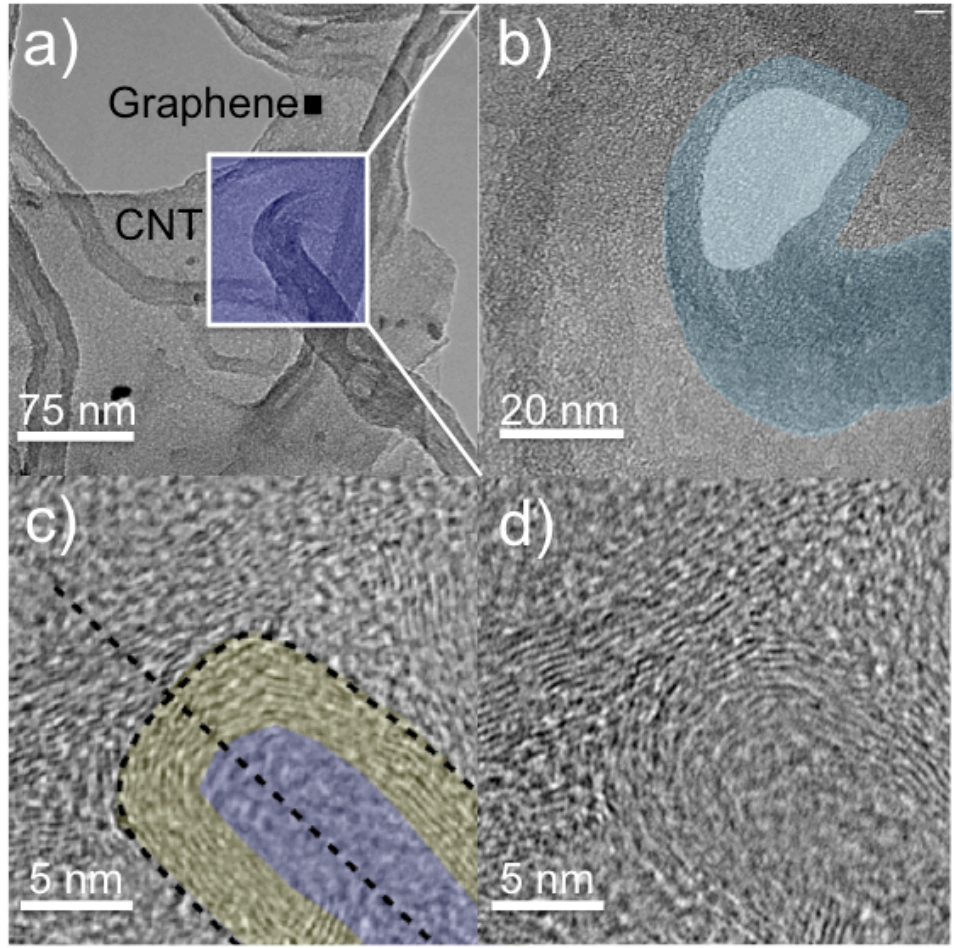

**Figure 1.** TEM images of the graphene-CNT hybrid. a) Low magnification TEM image of multiwalled CNTs growing out of the underlying graphene. The blue region highlights the root of the CNT originating from graphene. The average diameter for this CNT is 26.8±1.9 nm, b) Magnified, false colored image of the highlighted region in a). The blue color indicates the CNT walls extending from the graphene support and the white color further highlights the root region. c) High magnification, false colored TEM image of a different CNT atop graphene. The yellow and purple colors highlight the CNT walls and hollow inner tube, respectively. The graphene planes within the yellow region are parallel to the tube growth axis (black dashed trace), verifying that CNTs have been grown. The inner and outer tube diameters are 5.8 and 12.5 nm, respectively. d) Unmarked image of CNT in c).

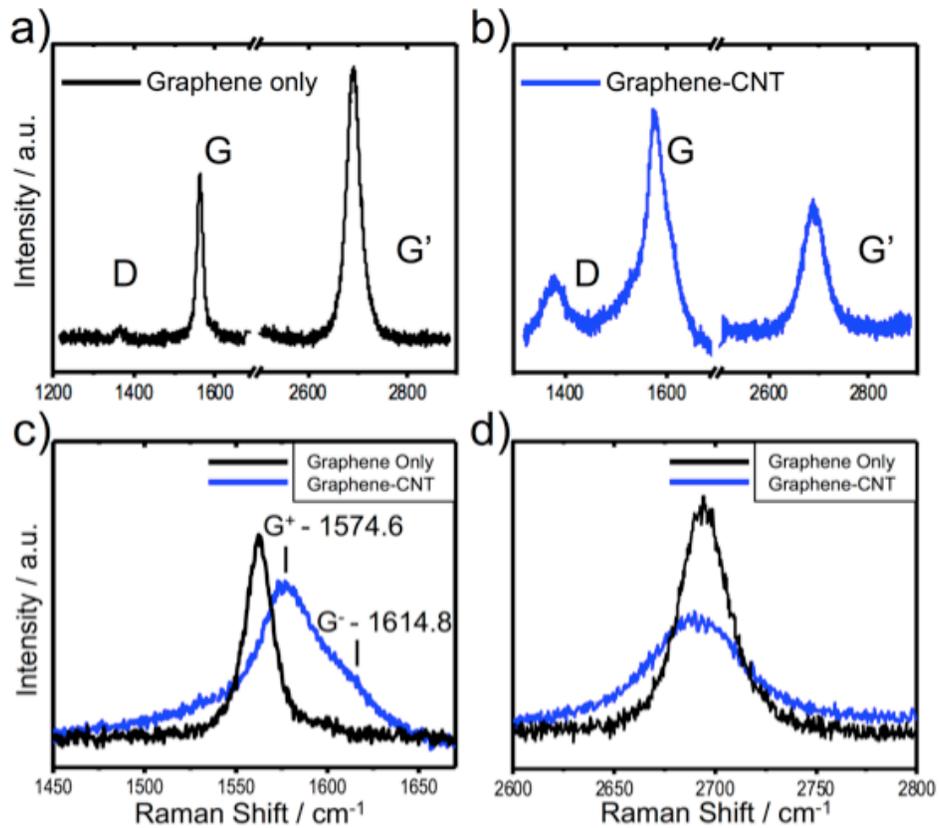

**Figure 2.** Raman spectra of a) CVD grown graphene and b) the graphene-CNT hybrid structure on SiO$_2$/Si substrate detailing broader bands and greater D band intensity for the graphene-CNT nanoarchitecture. c) Magnification of the G bands of both samples. While the G band of graphene is sharp and symmetric, the presence of CNTs in the graphene-CNT sample causes the G band to decrease in intensity and split into the G$^+$ and G$^-$ components. d) Magnification of the G' bands of both samples further exhibiting band broadening and significant intensity decrease due to CNT growth.

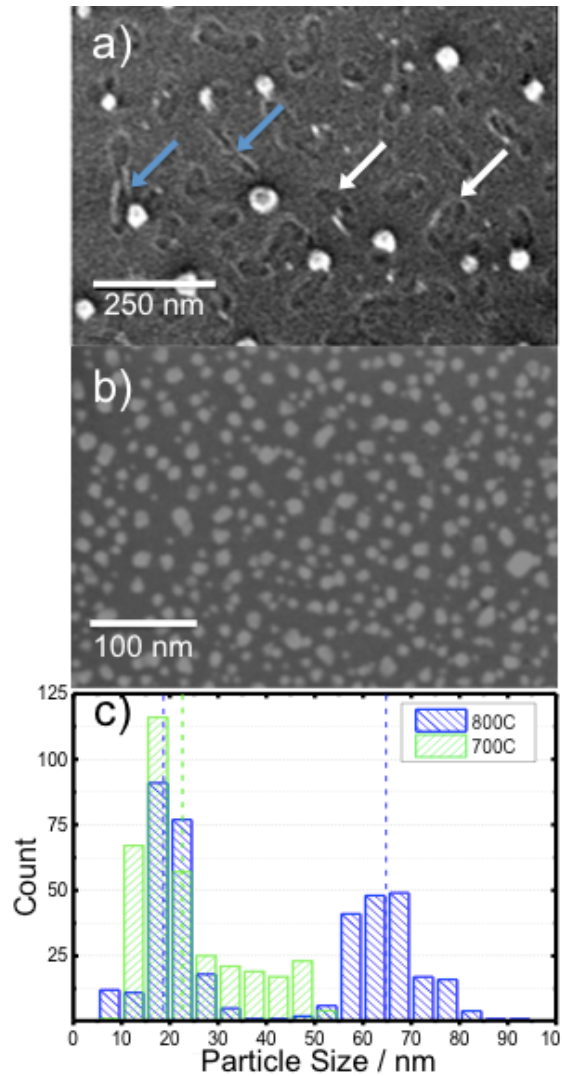

**Figure 3.** a), b) SEM images of Ni catalyst nanoparticles on a graphene substrate created via thermal treatment of Ni thin film at 800°C and 700°C, respectively. The white and blue arrows in a) identify etched sites due to stationary and mobile nanoparticles, respectively. c) Histogram of nanoparticle size. The 800°C growth displays a bimodal shape with two mean diameters at 19.3±5.3 nm and 64.5±7.9 nm. The 700°C is unimodal with mean diameter of 23.2±7.7 nm. The error values are 1 standard deviation from the mean.

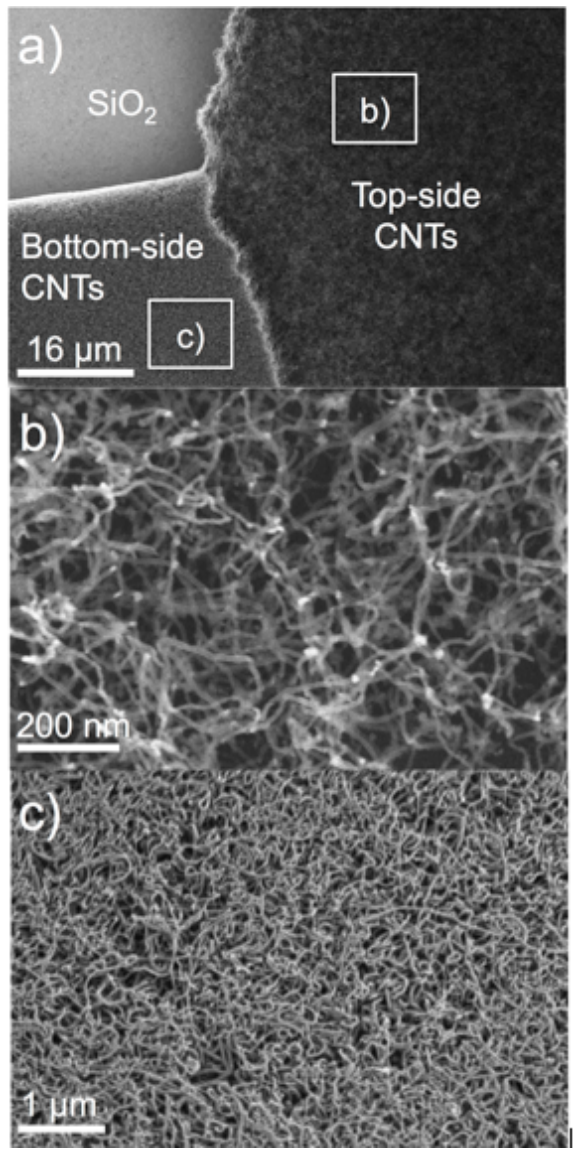

**Figure 4.** SEM images of CNTs grown on graphene at 800°C under gas flows Ar/H$_2$/C$_2$H$_4$ = 400/50/50 sccm. Here, Ni catalyst nanoparticles were created at 800°C. a) Top view of CNTs as-grown from catalysts on graphene. Sample is intentionally broken and peeled off for purpose of observation. b) Magnification of the top-side CNTs. c) Magnification of the bottom-side of the graphene-CNT sample with no graphene (fully etched away).

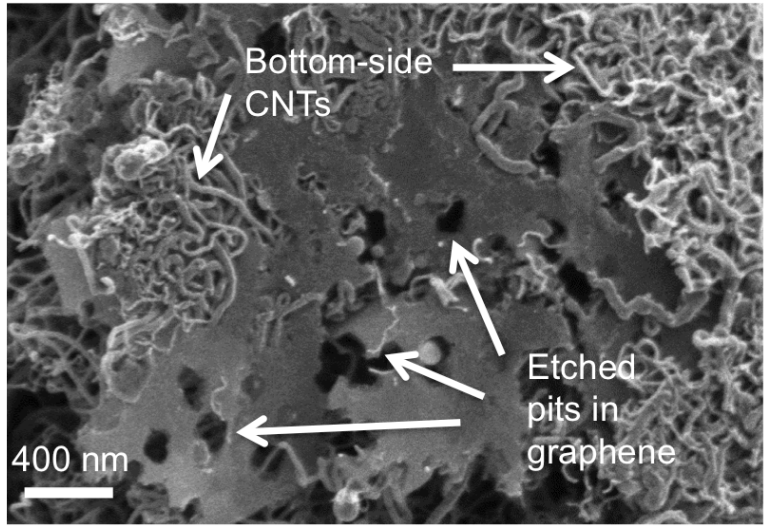

**Figure 5.** SEM image of the bottom-side of a graphene-CNT sample with partially etched graphene grown under gas flows Ar/H$_2$/C$_2$H$_4$ = 400/50/50 sccm. Here, Ni catalyst nanoparticles were generated at 700°C, whereas CNT growth with C$_2$H$_4$ occurred at 800°C.

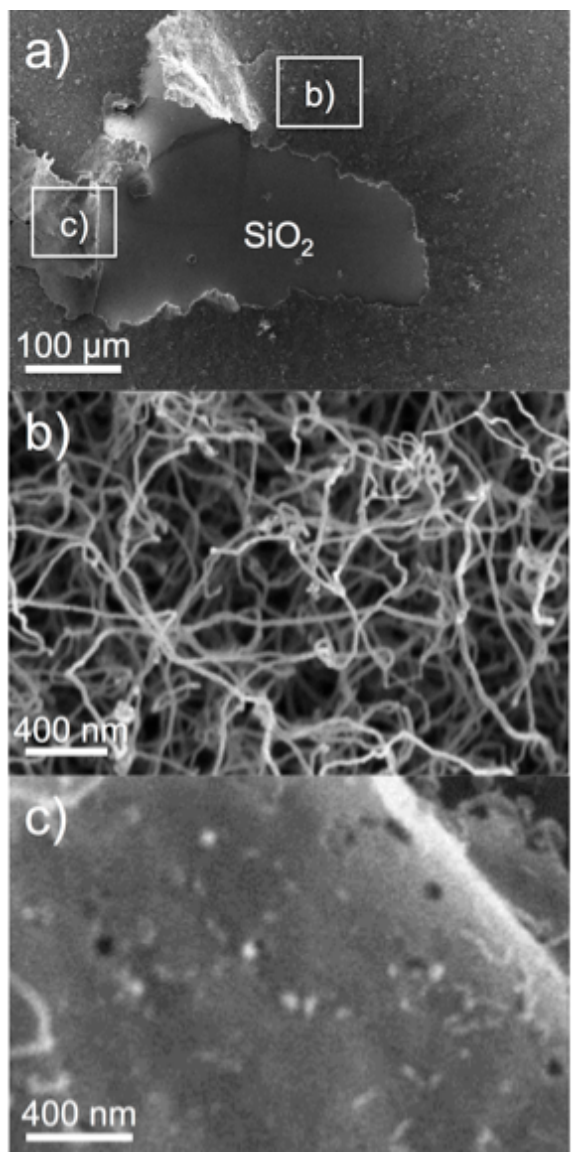

**Figure 6.** SEM images of CNTs directly grown on graphene at 700°C under gas flows Ar/H$_2$/C$_2$H$_4$ = 400/50/50 sccm. Here, Ni catalyst nanoparticles were formed at 700°C. a) Top view of CNTs as-grown on graphene. Sample is intentionally broken and peeled off for purpose of observation. b) Magnification of the top-side CNTs. c) Magnification of the bottom-side of the graphene-CNT sample with little to no graphene etching. The few CNTs observed in c) are a result of the mechanical scratching and peeling steps used to prepare the sample for SEM observation. They are not related to graphene etching.

# Supporting Information for

# Chemical vapor deposition of carbon nanotubes on monolayer graphene substrates: reduced etching via suppressed catalytic hydrogenation using $C_2H_4$

*Kitu Kumar[1], Youn-Su Kim[1], Xin Li[1], Junjun Ding[1], Frank T. Fisher[1], Eui-Hyeok Yang[1]\**

Department of Mechanical Engineering, Stevens Institute of Technology, Castle Point-on-Hudson, Hoboken, NJ 07030 USA

eyang@stevens.edu

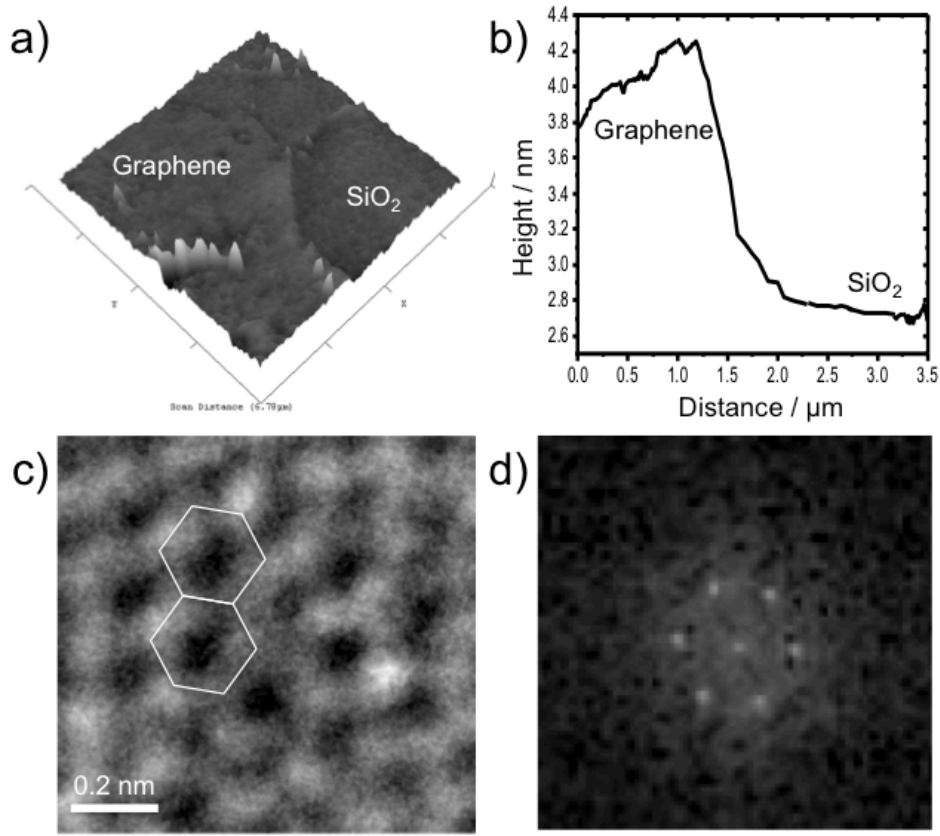

**Figure S1.** a) AFM image of graphene on $SiO_2$ displaying b) a step height of ~ 1nm. While the thickness of an atomic layer of graphite is reported to be 0.34 nm[1] the wet transfer process of CVD graphene traps a layer of water and impurities between graphene and substrate, thus increasing the measured thickness[2,3]. c) A high magnification, low-bandpass filtered image of a representative graphene lattice. Each bright spot represents the A-B atoms of graphene. White lines are a guide to the eye. d) Contrast-enhanced fast fourier transform (FFT) of c) detailing six-fold symmetry, which is a signature of monolayer graphene.

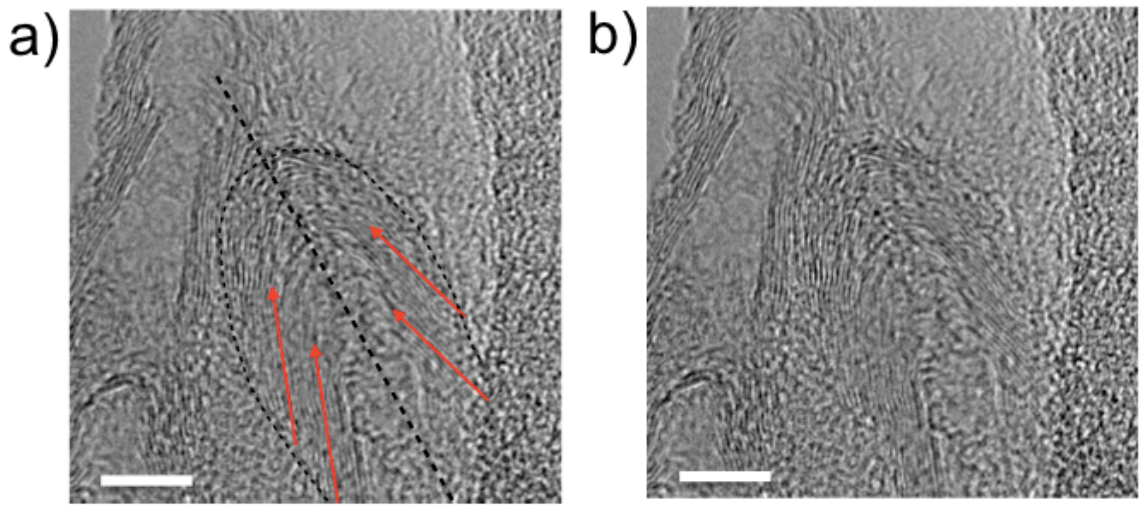

**Figure S2.** TEM images of a carbon nanofiber. In a) red arrows are parallel to the fishbone arrangement of graphene planes which are oriented at an angle to the carbon nanofiber growth axis (black dashed trace). In CNTs, the graphene planes are parallel to the growth axis. b) Unmarked image of carbon nanofiber in a). Scale bars are 5 nm.

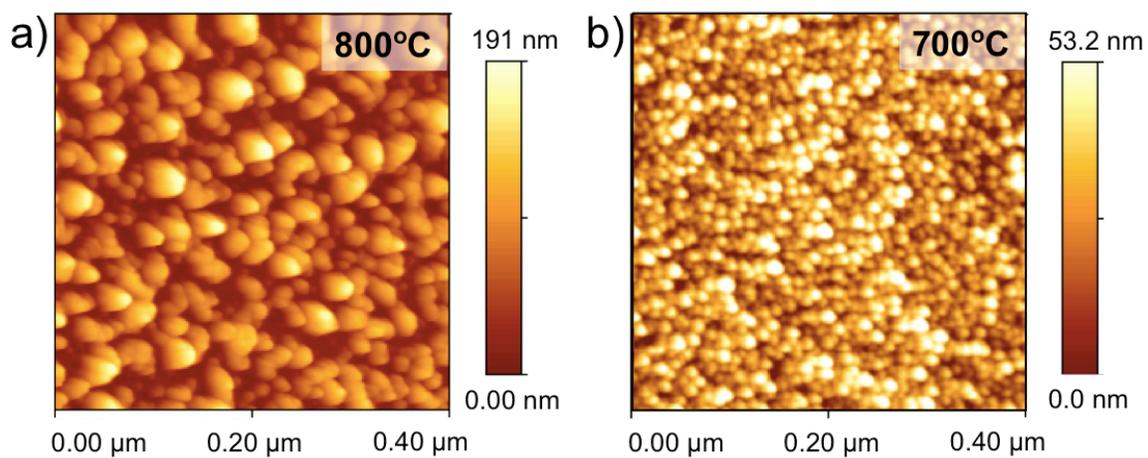

**Figure S3.** AFM Images of Ni catalyst nanoparticles on graphene via thermal treatment of a 3 nm Ni film at a) 800°C and b) 700°C. The mean circularity of the cross section of the nanoparticles in a) and b) is 0.611±0.304 and 0.894±0.163, respectively. The error value is 1 standard deviation from the mean. A circularity of 1.000 signifies a perfect circle.

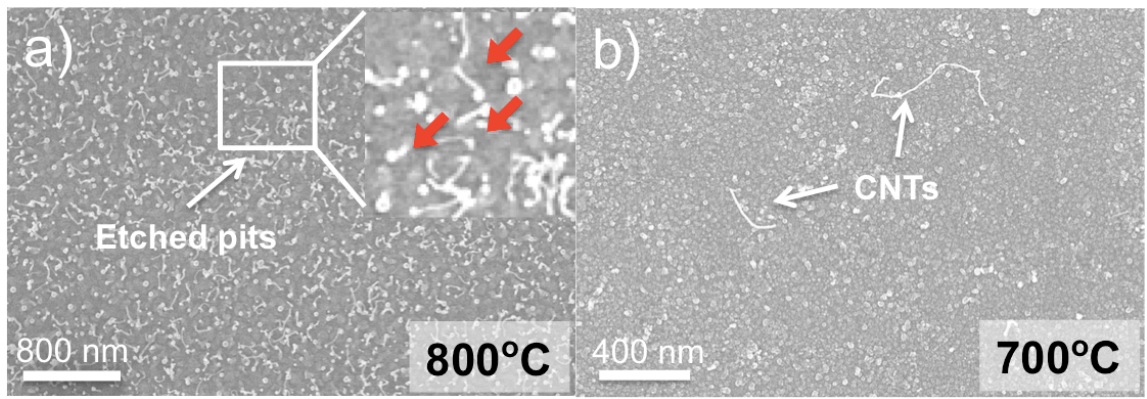

**Figure S4.** Effect of temperature on CNT growth on graphene substrates using $CH_4$. SEM images of CNT growth on a graphene substrate at a) 800°C and b) 700°C. The Ni catalyst nanoparticles were also generated at 800°C and 700°C, respectively. In a), more CNTs are grown and graphene etching is more pronounced (red arrows, inset) than in b). Thus, CNTs are grown more efficiently with $CH_4$ at higher temperatures at the expense of the graphene substrate. These results justify the use of $C_2H_4$ as in the manuscript where the higher hydrocarbon conversion ratio at 700°C permits greater carbon saturation in the Ni catalyst nanoparticles and suppresses the hydrogenation reaction that damages the graphene substrate.

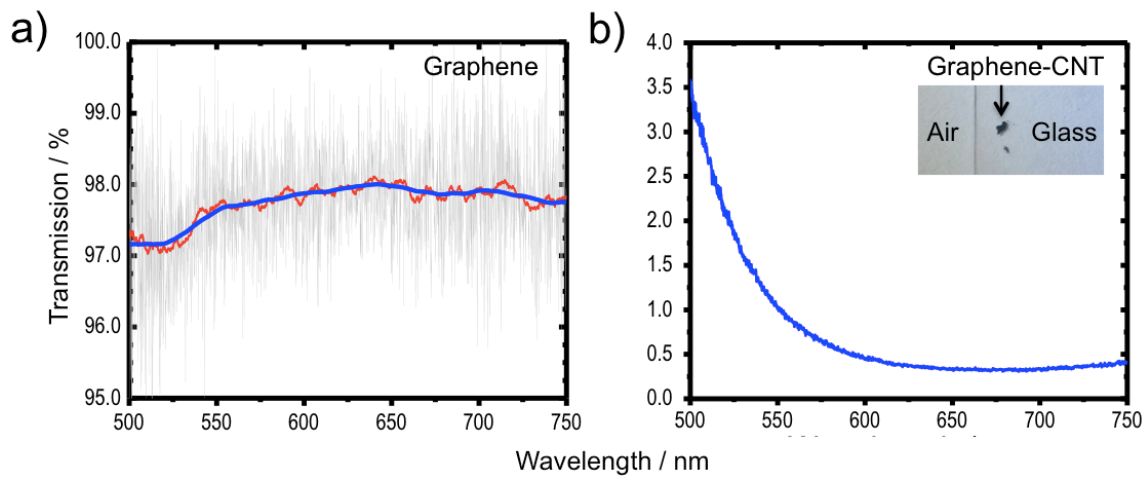

**Figure S5.** Transmission spectra as a function of incident wavelength on a) graphene and b) graphene-CNT samples. In a), the blue and red traces are smoothed fits to the original data (grey). Monolayer graphene absorbs 2.3% of light[4] and transmits 97.7%, therefore the sample in a) is predominantly monolayer. Inset to b), graphene-CNT samples are black, and should absorb most white light; light transmission is 0.5-3.0% in the same energy regime.

**Table S1** Extracted values from Raman spectra in **Figure 2** of the manuscript

| Peak Parameters | Energies (cm$^{-1}$) | | | FWHM (cm$^{-1}$) | | | Intensity Ratios | | Area Ratios | |
|---|---|---|---|---|---|---|---|---|---|---|
| Sample | D | G | G' | D | G | G' | $I_D/I_G$ | $I_{G'}/I_G$ | $I_D/I_G$ | $I_{G'}/I_G$ |
| Graphene | 1364.3 | 1562.6 | 2691.4 | 28.7 | 17.2 | 31.8 | 0.20 | 1.57 | 0.11 | 3.21 |
| Graphene-CNT | 1380.1 | 1574.6/ 1614.8* | 2690.7 | 68.0 | 39.8 | 53.9 | 0.50 | 0.73 | 0.58 | 0.90 |

*G$^+$/G$^-$ Energies

**Supporting Information References**